\documentstyle[prl,aps,floats,epsfig]{revtex} 

\begin{document}
\draft
\twocolumn[\hsize\textwidth\columnwidth\hsize\csname
@twocolumnfalse\endcsname

\title{Compositional Inversion Symmetry Breaking in Ferroelectric
Perovskites}

\author{Na Sai, B. Meyer, and David Vanderbilt }

\address{Department of Physics and Astronomy,\\
         Rutgers University, Piscataway, New Jersey 08854-8019, USA}
\date{March 14, 2000}
\maketitle

\begin{abstract}
Ternary cubic perovskite compounds of the form
(A$_{1/3}$A$'_{1/3}$A$''_{1/3}$)BO$_3$ and
A(B$_{1/3}$B$'_{1/3}$B$''_{1/3}$)O$_3$, in which the
differentiated cations form an alternating series of monolayers,
are studied using first-principles methods.  Such compounds are
representative of a possible new class of materials in which
ferroelectricity is perturbed by compositional breaking of
inversion symmetry.  For isovalent substitution on either
sublattice, the ferroelectric double-well potential is found to
persist, but becomes sufficiently asymmetric that minority
domains may no longer survive.  The strength of the symmetry
breaking is enormously stronger for heterovalent substitution, so
that the double-well behavior is completely destroyed.  Possible
means of tuning between these behaviors may allow for the
optimization of resulting materials properties.
\end{abstract}


\vskip2pc]
\baselineskip=11.6pt

\narrowtext

In the last decade, the extensive use of first-principles theoretical
methods to study ferroelectric perovskite oxides has led to
a greatly expanded understanding of the ferroelectric and
piezoelectric properties of this important class of
materials.  Theoretical investigations of electronic, dynamical, and
structural properties have been shown
to be in good accord with experimental observations
for the simple ABO$_3$ perovskites \cite{Opinion}.
Moreover, these studies provide microscopic insight into the
ferroelectric instabilities, their relation to the long-range
Coulomb interactions\cite{Ghosez}, and related questions about the origins
of the piezoelectric response\cite{Szabo,LBDVpiezo}.

However, the materials of most interest for technological
applications are generally not the simple ABO$_3$ perovskites, but
{\it solid solutions} with stoichiometric substitutions of A or B metal atoms.
Examples include PZT (PbZr$_{1-x}$Ti$_x$O$_3$), currently one of the most
widely used ferroelectrics, and PZN (PbZn$_{1/3}$Nb$_{2/3}$O$_3$)
and PMN (PbMg$_{1/3}$Nb$_{2/3}$O$_3$) and their solid solutions
with PbTiO$_3$, which have recently been shown to have enormous
piezoelectric response in single-crystal form \cite{Park}.
In fact, this class of materials shows great promise for the
development of {\it new materials} having improved dielectric
and electromechanical properties.  Not only is there an enormous
space of chemical compositions to explore, but it may also be
possible to optimize the desired material properties by tuning the
degree and type of compositional order for the desired
application \cite{Setter,Davies}.

In particular, Eckstein \cite{Eckstein} has recently
suggested that the artificial atomic-layer growth of {\it
compositionally ordered structures that break inversion symmetry}
might be especially exciting and fruitful in this regard.
The resulting asymmetry of the ferroelectric double-well potential in
such a material suggests the prospect of {\it qualitatively new behavior,}
e.g., ``self-poling'' materials with tailored piezoelectric or
dielectric properties.

\begin{figure}
\centerline{\epsfig{file=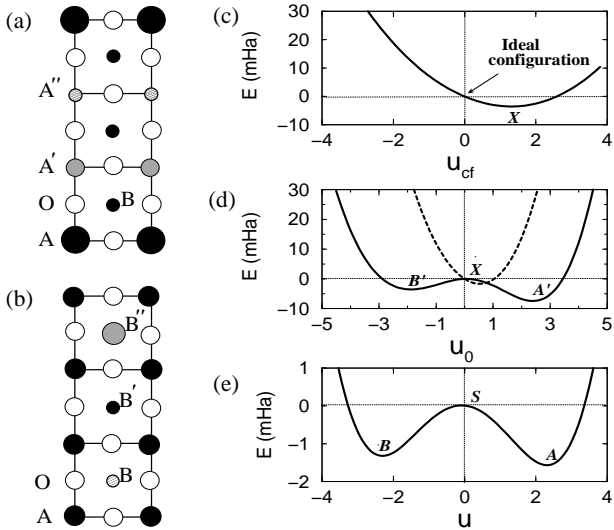,height=7cm}}
\vskip 5pt
\caption{Side views of
(a) (A$_{1/3}$A$'_{1/3}$A$''_{1/3}$)BO$_3$, and
(b) A(B$_{1/3}$B$'_{1/3}$B$''_{1/3}$)O$_3$ structures.
Energy {\it vs.} displacement along
(c) the cubic force direction,
(d) the line ${\bf r}={\bf r}_X+u_0\hat{\xi}_0$, and
(e) the line connecting the minima $A$ and $B$.
Solid (dashed) curves are for A-iso (B-hetero) systems.
\label{fig:procedure}}
\end{figure}

In this Letter, we explore this exciting prospect by carrying
out {\it ab-initio} theoretical calculations for several prospective
model structures of this type.  Specifically, we envision the
artificial growth of materials of overall composition
(A$_{1/3}$A$'_{1/3}$A$''_{1/3}$)BO$_3$ or
A(B$_{1/3}$B$'_{1/3}$B$''_{1/3}$)O$_3$,
in which the three different cations alternate layer-by-layer
along the ferroelectric direction
as illustrated in Figs.~\ref{fig:procedure}(a-b) for A-site and B-site
modulation respectively.  We will {\it assume} that
such atomic-layer control will become possible \cite{Eckstein},
and investigate the energy landscapes and ferroelectric
properties of the resulting materials.
As will be shown below, we find that the asymmetry may easily be strong
enough for self-poling to occur, even in the case of isovalent
substitution.  Moreover, we find a surprisingly strong
variation of the strength of the symmetry breaking with the strength of
the compositional perturbation, allowing for a very wide tunability of
ferroelectric properties.  Experimentally, such tunability might be
exploited by growing alloy structures with a cyclic modulation of the
concentration variable.
We thus find very strong motivations for the development of such
materials, and hope that our work will encourage experimental efforts
directed toward their synthesis.


We adopt bulk BaTiO$_3$ as a prototypical parent compound
and construct a series of model systems that allow us to test
for the effects of A-site {\it vs.}\ B-site and
isovalent {\it vs.}\ heterovalent substitution.  Specifically,
we consider (Ba$_{1/3}$Sr$_{1/3}$Ca$_{1/3}$)TiO$_3$ as an example
of an ``A-iso'' system (isovalent substitution on the A site),
Ba(Sc$_{1/3}$Ti$_{1/3}$Nb$_{1/3}$)O$_3$ as an example of a
``B-hetero'' system (heterovalent substitution on the B site), and
Ba(Ti$_{1/3}$Zr$_{1/3}$Hf$_{1/3}$)O$_3$ as an intermediate
``B-iso'' case.
We construct 15-atom supercells by tripling the primitive unit
cell along the $z$ direction and cyclically alternating the identity
of the A or B atom, as shown in Figs.~1(a-b).  We assume that any
ferroelectric order develops only along the $z$-direction, so that
the material remains tetragonal,
and only $z$ displacements need be considered.
(It may be possible to realize this situation by
appropriate choice of lattice-mismatched substrate for epitaxial
growth, but it is not our purpose here to investigate this
possibility.)  We also assume perfect control
of layer-by-layer composition, resulting in the ideal stacking of
Figs.~1(a-b), and carry out our theoretical studies only at
zero temperature.  Much future work clearly remains to be done in
relaxing these assumptions, but the study of simple prototypical
systems is a natural first step.


The {\it ab-initio} calculations are carried out using the
Vanderbilt ultra-soft pseudopotential scheme\cite{David90}
in the local-density approximation. Details of the
pseudopotentials can be found in Ref.~\onlinecite{KSDV94}. Good $k$-point
convergence is obtained using a (6,6,2) Monkhorst-Pack mesh,
corresponding to the bulk (6,6,6) mesh\cite{explan-k}, and a 25-Ry plane-wave cutoff is used throughout.
Because experimental lattice constants $a$ and $c$ are not available
for the ordered compounds of interest, the calculated theoretical
equilibrium values were usually used.  However, in some cases the
$c/a$ ratio was fixed and only the cell volume was optimized, as
detailed below.

In the presence of the broken inversion symmetry, it is not always
easy to locate both local energy minima corresponding to the two
ferroelectric ground states.  We have found that the following
procedure is quite reliable for finding both minima, if they exist.
First, we place the atoms at the ideal cubic coordinates and calculate
the pattern of forces, thus defining a ``cubic force direction''
$\hat{\xi}_{\rm cf}$ in the 15-dimensional configuration space.
We then use $\hat{\xi}_{\rm cf}$ as the search direction for a line
minimization, and denote the minimum along this line
${\bf r}=u_{\rm cf}\,\hat{\xi}_{\rm cf}$ to be $X$, as
shown in Fig.~\ref{fig:procedure}(c).  Next, starting from $X$,
we carry out a second line minimization along the line
${\bf r}={\bf r}_X+u_0\,\hat{\xi}_0$, where
$\hat{\xi}_0$ corresponds to the ferroelectric mode unit vector of
bulk BaTiO$_3$.  As shown in Fig.~\ref{fig:procedure}(d), two scenarios
can be identified.  (i) If two minima $A'$ and $B'$ are
found along this line, then we carry out a steepest-descent
minimization from each, obtaining two distinct local minima $A$ and $B$.
Fig.~\ref{fig:procedure}(e) shows the perturbed double-well potential
plotted along a line connecting points $A$ and $B$ (direction
$\hat{\xi}$) in such a case.
(ii) If only one minimum is found, as for the dashed line of
Fig.~\ref{fig:procedure}(d), then the double-well potential has
been destroyed, and a subsequent steepest-descent minimization
identifies the unique minimum.


\begin{figure}
\centerline{\epsfig{file=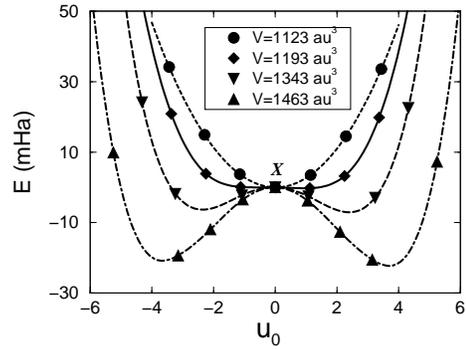,angle=270,width=6cm}}
\vskip 5pt
\caption{Energy vs. displacement along the line
${\bf r}={\bf r}_X+u_0\,\hat{\xi}_0$ in the A-iso system for
several cell volumes.
\label{fig:Aiso}}
\end{figure}

We first consider the application of this procedure to the A-iso system
(Ba$_{1/3}$Sr$_{1/3}$Ca$_{1/3}$)TiO$_3$. The relaxation from the
ideal cubic structure to the point $X$ illustrated in
Fig.~\ref{fig:procedure}(c) is found to be dominated by ionic
size effects:  one sees a simple shift of the atoms toward
smaller A-site cations and away from larger ones.
Figure \ref{fig:procedure}(d) then illustrates the
minimization along ${\bf r}={\bf r}_X+u_0\,\hat{\xi}_0$ for one
particular fixed cell volume (1343\,a.u.$^3$),
indicating a modest
asymmetry between minima $A'$ and $B'$.  However, since it is well
established that the presence or absence of a ferroelectric
instability depends strongly on cell volume in perovskites
\cite{KSDV94}, we have repeated this step for a series of fixed
atomic volumes in Fig.~\ref{fig:Aiso}.
(The curves in Fig.~\ref{fig:Aiso} could thus
be roughly interpreted as corresponding
to a series of related compounds having differing tendencies
towards ferroelectric instability.)  The results illustrate the
typical evolution of the double-well potential with increased
ferroelectric tendency, from a slightly asymmetric single-well
minimum ($V$=1123\,a.u.$^3$) to a slightly asymmetric double-well
potential ($V$=1463\,a.u.$^3$) \cite{explan-ca}.
It happens that the theoretical
equilibrium volume $V$=1193\,a.u.$^3$ is a nearly borderline case;
we find only a single minimum, but there are still two inflection
points.  In such a case, we modify the procedure of the previous
paragraph by carrying out a steepest-descent minimization from
each inflection point, and by doing so we succeed in finding the
two distinct minima ${\bf r}_A$ and ${\bf r}_B$ in
Fig.~\ref{fig:procedure}(e).  For this case ($V$=1193\,a.u.$^3$),
we find a soft-mode amplitude of 0.3\,a.u.\ and an average well depth
(relative to the saddle point $S$) of 0.48\,mHa per 5-atom cell,
fairly close to the corresponding values of 0.25\,a.u.\ and 0.43 mHa
\cite{KSDV94}, respectively, for tetragonal bulk BaTiO$_3$.
The energy difference between the twin wells is modest,
$\sim 15-20\%$ of the well depth.

Applying the same approach to the B-iso system
Ba(Ti$_{1/3}$Zr$_{1/3}$Hf$_{1/3}$)O$_3$, we
observe very similar behavior: a strong volume-dependence of
the ferroelectric tendency, and a modest inversion asymmetry
for both single-well and double-well volumes. In this case,
the structure does not develop a double-well potential until
the volume is increased to 1468\,a.u.$^3$,
but the degree of asymmetry is similar as for the A-iso case.
Thus, we conclude that the choice of A vs.~B site for an isovalent
chemical substitution does not strongly affect the strength of
the asymmetry or the qualitative behavior of the system.

While the asymmetries may appear small in the case of isovalent
substitution, they are large by one important measure.  We define
an ``effective electric field'' ${\cal E}_{\rm eff}=\Delta E/\Delta P_s$,
where $\Delta E$ and $\Delta P_s$ denote the energy and polarization
differences between the two local minima.  If this
quantity is larger than the coercive field of the material, it
means that the thermodynamic preference for the energetically
preferred minimum is strong enough to overcome the pinning of the
domain walls and to spontaneously switch the material into a
single-domain state.  Using the Berry-phase approach \cite{KSDVBerry}
to calculate $\Delta P_s$, we find ${\cal E}_{\rm eff}$=90\,kV/cm at
the equilibrium volume
in the A-iso system.  Since the typical coercive field of most
perovskite ferroelectrics is closer to 15\,kV/cm, we thus arrive at
the important conclusion that the symmetry breaking may easily be
strong enough to cause the material to self-pole, even in the
case of isovalent substitution.

Turning now to the case of {\it heterovalent} chemical substitution,
we find very different behavior in this case.  Specifically, we
consider the B-hetero system
Ba(Sc$_{1/3}$Ti$_{1/3}$Nb$_{1/3}$)O$_3$ in which the valence
charges are +3, +4, and +5 on Sc, Ti, and Nb, respectively.
Structural optimization results in a lattice constant of
$a=7.60$\,a.u. and $c/a=3.023$.  The pattern of relaxation
$\hat{\xi}_{\rm cf}$ leading to configuration $X$ suggests that
the electrostatic interaction is a dominant effect.
Specifically, we observe bucklings of the AO and BO$_2$ planes
that are consistent with a picture of static electric fields
arising from the different B-atom valence charges.

Most importantly, for the heterovalent case we find {\it only a
single minimum} when searching from $X$ along $\hat{\xi}_0$, as
shown by the dashed curve in Fig.~\ref{fig:procedure}(d).  Various
alternate search strategies failed to identify a second minimum;
all trial structures were found to relax back to a unique
structural ground state.  The absence of the second minimum was not
found to be sensitive to the cell volume, as was the case for
isovalent substitutions.  Thus, we find that the symmetry breaking is
enormously stronger than for the isovalent case, and it is clear that
resulting behavior is of a qualitatively different type.

\begin{figure}
\centerline{\epsfig{file=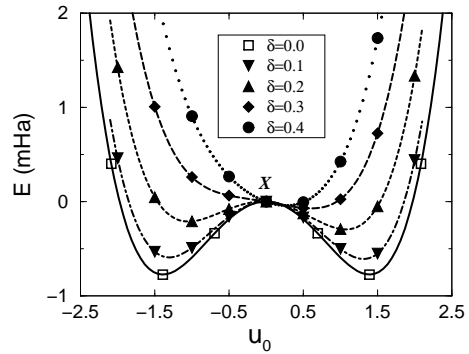,angle=270,width=6cm}}
\vskip 5pt
\caption{Energy vs. displacement along the line ${\bf r}={\bf
r}_X+u_0\,\hat{\xi}_0$ in the B-hetero system with continuous
$\delta$.
\label{fig:B_het}}
\end{figure}

In order to gain a better understanding of this behavior, and in
particular to track the disappearance of the secondary minimum,
we have developed a model system to study
the effects of ``turning on'' the heterovalent symmetry-breaking
perturbation gradually.  We introduce a continuous variable
$\delta$, and construct artificial atoms with fractional nuclear
charges that deviate by $\pm\delta$ from those of Ti ($Z$=22).
Constructing a crystal out of a cyclic series of Ti$-\delta$, Ti,
and Ti$+\delta$ atomic layers, we can continuously tune the system
from ferroelectric BaTiO$_3$ ($\delta$=0) to a full-fledged heterovalent
system Ba(Sc$_{1/3}$Ti$_{1/3}$V$_{1/3}$)O$_3$ ($\delta$=1).  The
energies calculated along the $\hat{\xi}_0$ direction from $X$ are
plotted for a series of $\delta$ values in Fig.~\ref{fig:B_het}.
As $\delta$ increases from 0 to 0.3, the asymmetry of the double
well increases and the well depth decreases.  At $\delta=0.4$, the
curve exhibits only a single energy minimum, signaling the
transition from double-well to single-well behavior in response to
the increasingly strong symmetry-breaking perturbation.
To confirm the disappearance of the secondary minimum more directly,
we also tracked its evolution as $\delta$ was ``turned on'' in a
sequence of small steps, using the relaxed structure at one $\delta$
as a starting guess for the next.
At $\delta=0.34$,
the minimum was confirmed to disappear, and subsequent relaxation
led back to the principal (now global) minimum.

To gain more insight into the structural relaxations, we have found
it useful to introduce a measure of the
``strength of the symmetry breaking.''  While it is tempting to
choose a measure that is related to the energy difference between
the two local minima, like the ${\cal E}_{\rm eff}$ introduced
earlier, such a definition has the disadvantage of being ill-defined
in the single-minimum case.  Thus, we have adopted instead the
following measure.  For any
curve such as that of Figs.~1(d-e), 2, or 3, we locate the point
of minimum $d^2E/du^2$ (i.e., vanishing $d^3E/du^3$), and then
define $F_{\rm sb}$ to be $dE/du$ evaluated at that point.
We refer to $F_{\rm sb}$ as the ``symmetry-breaking force''
since it measures the strength of the symmetry breaking and has
units of force.  It turns out to have a similar behavior
as the well-depth difference in the double-well case, but has
the advantage of remaining well-defined in the single-well case.

Calculations of $F_{\rm sb}$ in the A-iso and B-iso
cases indicate that $F_{\rm sb}$ has only a modest and smooth
dependence on cell volume even when passing through the
transition from single-well to double-well behavior, confirming
that the ``strength of the symmetry breaking'' is not the
variable parameter in those cases.  However, returning
to the virtual-atom B-hetero system, we find that $F_{\rm sb}$ is
an extremely sensitive function of $\delta$, with numerical fits
indicating a $\delta^3$ dependence.  It is not hard
to see that $F_{\rm sb}$ must be an odd function of $\delta$, but its
cubic behavior may seem surprising at first sight.  However, the
vanishing of the linear term can be deduced from simple symmetry
arguments.  An intuitive form of the argument is to note that
the symmetry-breaking perturbation, which is of the form
($+\delta$, $-\delta$, 0) in successive layers, can be regarded as
a superposition of two perturbations
($2\delta/3$, $-\delta/3$, $-\delta/3$) and
($\delta/3$, $-2\delta/3$, $\delta/3$) that {\it do not break}
inversion symmetry.  Thus, the principle of superposition prevents
the occurrence of any symmetry-breaking response in linear order
in $\delta$, and in particular $F_{\rm sb}$ must vanish.  (A more
systematic analysis may be made by considering the $C_{3v}$
symmetry group consisting of primitive translations
along $\hat{z}$ and mirrors $M_z$.  The ferroelectric mode vector
and the perturbation $\delta$ are found to belong to the $A_2$ and
$E$ respresentations, repsectively, and thus
cannot couple at linear order.)

Knowing the form of this extraordinarily strong cubic dependence of
$F_{\rm sb}$ on $\delta$, we can now understand the pronounced qualitative
differences that were observed for the cases of isovalent and
heterovalent substitution.  We find that $F_{\rm sb}$ is about the same,
$\sim$0.25\,mHa/a.u., in the B-hetero system with $\delta=0.25$
as in the A-iso system.  Increasing $\delta$ from 0.25 to 0.4 increases
$F_{\rm sb}$ by about a factor of four, enough to cause the transition to
single-well behavior.  A further increase of $\delta$ from 0.4 to
1.0 leads to a further increase of the symmetry-breaking force
$F_{\rm sb}$ by a {\it factor of about 15}.  Thus, in a fully-developed
B-hetero system such as Ba(Sc$_{1/3}$Ti$_{1/3}$Nb$_{1/3}$)O$_3$, 
the strength of the symmetry breaking is more than an order of
magnitude larger than needed to destroy the secondary minimum
completely.  It is hardly surprising, then, that we observed no
secondary minimum in this case!

The enormous disparity between the behavior in the isovalent and
heterovalent cases suggests that it may be of great interest to
find a way of tuning the system continuously from one behavior to
the other.  One can imagine doing this by regarding $\delta$ not
as a variable atomic number, but rather as representing a
layer-by-layer composition variable.  For example,
one could conceive of the epitaxial growth
of Ba(Sc$_{1-y}$Nb$_y$)O$_3$
in an alternating sequence of layers with $y=0.5(1+\delta)$, $y=0.5$,
and $y=0.5(1-\delta)$. By the same symmetry arguments, the effective
strength of the symmetry breaking must again scale as $\delta^3$.  Thus, by
controlling the concentration variable $\delta$, one can hope to tune
the system over a very wide range of behavior.

One would clearly like to use this tunability to optimize the
desired characteristics of the material, such as the piezoelectric
response.  It might naively be expected that increasing
the strength of the inversion symmetry breaking will increase the
piezoelectric response.  (For example, if the parent material at
$\delta$=0 were {\it paraelectric}, then the piezoelectric
response would be expected to scale as $\delta^3$.)  On the other
hand, the materials that have the largest piezoelectric
coefficients are typically ferroelectrics, and since a very
strong symmetry breaking suppresses the ferroelectric behavior,
it might be counterproductive to make $\delta$ too large.
Clearly, further theoretical investigation is needed in order to
clarify these issues.

Many other questions remain open and need to be resolved.  For
example, the conditions under which the polarization will remain
oriented parallel to the growth direction, the behavior of the
system as a function of temperature, and the properties of
materials with simultaneous A-site and B-site substitution,
are obvious candidates for further study.  The thermodynamic
behavior of these materials, which are technically pyroelectric
but interpolate to a ferroelectric limit as $\delta\rightarrow0$,
are also deserving of investigation.   In the meantime,
we hope that our theoretical investigations will stimulate
attempts at experimental growth and characterization of novel
perovskites with compositionally broken inversion symmetry.


We thank J.~Eckstein for suggesting the direction of this study.
Support for this work was provided by ONR Grant N00014-97-1-0048 and
NSF Grant DMR-9981193.  We thank M.~Cohen and K.~Rabe for useful
discussions.



\begin{references}

\bibitem{Opinion}D. Vanderbilt, Curr. Opin. Mater. Sci
{\bf 2}, 701 (1997); and J. Korean Phys. Soc. S103 (1998).

\bibitem{Ghosez} Ph. Ghosez, X. Gonze and J.-P. Michenaud,
Europhys. Letters {\bf 33}, 713 (1996); Ferroelectrics {\bf 186}, 73
(1996).

\bibitem{Szabo} G. Saghi-Szabo, R.E. Cohen and H. Krakauer,
Phys. Rev. Lett. {\bf 80}, 4321 (1998);  Phys. Rev. B {\bf 59},
12771 (1999).

\bibitem{LBDVpiezo} L. Bellaiche and David Vanderbilt, Phys. Rev.
Lett. {\bf 83}, 1347 (1999).

\bibitem{Park} S.-E. Park and T.E. Shrout, J. Appl. Phys. {\bf 82},
1804 (1997).

\bibitem{Setter} N. Setter and L.E. Cross, J. Appl. Phys. {\bf 51},
4356 (1980).

\bibitem{Davies} M.A. Akbas and P.K. Davies, J. Am. Ceram. Soc.
{\bf 81}, 670 (1998).

\bibitem{Eckstein} J. Eckstein, private communication.

\bibitem{David90}D. Vanderbilt, Phys. Rev. B {\bf 41},7892 (1990).

\bibitem{KSDV94} R.D. King-Smith and D. Vanderbilt, Phys. Rev. B
{\bf 49}, 5828 (1994).

\bibitem{explan-k} A (4,4,2) $k$-point grid was found to be
sufficient for the B-iso case.

\bibitem{explan-ca} These calculations were done at a fixed
$c/a=3.04$ derived from the bulk tetragonal phase of BaTiO$_3$.
See J. Padilla and D. Vanderbilt, Phys. Rev. B {\bf 56}, 1625
(1997).

\bibitem{KSDVBerry}R.D. King-Smith and D. Vanderbilt, Phys. Rev. B
{\bf 47},1651(1993).

\end{references}
\end{document}